\documentclass[prd,reprint,superscriptaddress]{revtex4-1}

\usepackage{latexsym,amssymb,amstext,amsmath}
\usepackage{slashed}
\usepackage{mathrsfs}
\usepackage{color}
\usepackage{hyperref}
\usepackage{verbatim}
\usepackage{graphicx}
\usepackage{color}
\usepackage{mathrsfs}

\def\rmi{{\rm i}}

\newcommand {\cL}{{\cal L}}

\newcommand {\cN}{{\cal N}}

\newcommand {\cW}{{\cal W}}

\def\a{\alpha}
\def\b{\beta}

\def\f{\phi}
\def\g{\gamma}

\def\l{\lambda}
\def\m{\mu}
\def\n{\nu}
\def\o{\omega}
\def\p{\psi}

\def\x{\xi}

\def\O{\Omega}

\newcommand{\nn}{\nonumber}

\newcommand{\vf}{\varphi}

\newcommand{\be}{\begin{equation}}
\newcommand{\ee}{\end{equation}}
\newcommand{\bea}{\begin{eqnarray}}
\newcommand{\eea}{\end{eqnarray}}

\newcommand{\ba}{\begin{array}}
	\newcommand{\ea}{\end{array}}

\def\double #1{#1{\hbox{\kern-2pt $#1$}}}

\newcommand{\bsubeq}{\begin{subequations}}
	\newcommand{\esubeq}{\end{subequations}}

\begin{document}
	
	\title{Superconformal Generalizations of  Auxiliary Vector Modified Polynomial $f(R)$ Theories}
	
	\author{Sibel Boran}
	\email{borans@itu.edu.tr}
	\affiliation{Department of Physics,
		Istanbul Technical University,
		Maslak 34469 Istanbul,
		Turkey}
	
	\author{Emre Onur Kahya}
	\email{eokahya@itu.edu.tr}
	\affiliation{Department of Physics,
		Istanbul Technical University,
		Maslak 34469 Istanbul,
		Turkey}
	
	\author{Nese Ozdemir}
	\email{nozdemir@itu.edu.tr}
	\affiliation{Department of Physics,
		Istanbul Technical University,
		Maslak 34469 Istanbul,
		Turkey}
	
	\author{Mehmet Ozkan}
	\email{ozkanmehm@itu.edu.tr}
	\affiliation{Department of Physics,
		Istanbul Technical University,
		Maslak 34469 Istanbul,
		Turkey}
	
	\author{Utku Zorba}
	\email{zorba@itu.edu.tr}
	\affiliation{Department of Physics,
		Istanbul Technical University,
		Maslak 34469 Istanbul,
		Turkey}
	
	

	\begin{abstract}
		
		We present the supersymmetric completion of the auxiliary vector modified polynomial $f(R)$ theories in their dual scalar-tensor theory formulation that interpolate between the auxiliary vector modified polynomial $f(R)$ theories and chaotic inflation with the power-law potential $V(\f) \propto \f^p$. The supersymmetrization is achieved in two steps: First, we introduce a superconformal theory for three chiral multiplets by choosing a conformal K\"ahler potential and a conformal superpotential. In the second step, we use one of the chiral multiplets to compensate for the superconformal symmetries and achieve the K\"ahler potential and the superpotential while the other two are used to realize inflation with a stable inflationary trajectory. The stability of the inflationary trajectory requires certain deformations to the K\"ahler potential which we discuss their compatibility against the inflationary observables from the latest Planck data.
		
	\end{abstract}
	
	
	\maketitle
	\allowdisplaybreaks
	\section{Introduction}
	After Planck Collaboration had released the data collected by the Planck satellite in 2013  \cite{Planck2013}, there has been an increasing number of attempts to construct inflationary models that are ranging from scalar-tensor theories to higher-order curvature invariants (for reviews, see e.g. \cite{DeFelicefR,SotirioufR,Encyc}) and test their predictions against the inflationary observables. 
	In particular, the observational constraints on the spectral tilt $n_s$ and tensor to scalar ratio $r$ severely restricts the spectrum of possible inflationary models but they are not decisive in the sense that many inflationary models become identical during inflation. Consequently, they predict the same inflationary observables up to the next order contribution to the inflationary parameters\cite{Kehagias}. This universality of a large class inflationary models begs for a theoretical argument to clarify why different theories are almost indistinguishable during inflation.
	
	In Ref.\cite{Nonmin}, a guiding principle for the universality was given from a perspective of (super)conformal symmetry breaking. There, one first considers the maximal extension of the (super)Poincar\'e group that is the (super)conformal group. The addition of the extra symmetries leads uniquely to a particular scalar-tensor theory whose gauge fixing gives rise to a pure de Sitter (or anti-de Sitter) solution depending on the sign of the cosmological constant.  When inflation is realized within that paradigm, one naturally needs at least two scalar fields, one being the conformon that is to be used as a compensator for the conformal symmetries and the other being the inflaton. As noted in \cite{universality}, if there is a $SO(1,1)$ symmetry between inflaton and conformon fields, then it is not possible to realize a scalar potential but a pure de Sitter (or anti-de Sitter) solution still insists. Therefore, a smooth deviation from a de Sitter space translates into a smooth deformation of the $SO(1,1)$ symmetry between the scalar fields. This line of thinking lead one to T-model of inflation with the scalar potential 
	\bea
	V(\vf) = \l_n \tanh^{2n} \frac{\vf}{\sqrt 6}\,,
	\eea
	that is insensitive to the power $n$ that plays the role of deviation from $SO(1,1)$ symmetry. This model has an attractor point 
	\bea
	1- n_s = \frac{2}{N}\,, \qquad r = \frac{12}{N^2} \,,
	\label{Attractor}
	\eea
	with $N$ being the number of e-folds between horizon exit and the end of inflation.These values of $n_s$ and $r$ perfecty fit to the latest Planck data \cite{Planck2018}. For the future reference, this inflationary predicions coincide with the predictions of the Starobinsky model that is described by the Lagrangian \cite{Starobinsky}
	\bea
	e^{-1} \cL = R  + \frac{1}{6M^2} R^2 \,,
	\eea
	which belongs to the class of conformal chaotic inflationary models \cite{SuperStraobinsky}.  This consideration was later generalized to $\a$-attractor T-Model inflation with the scalar potential that is given by
	\bea
	V(\vf) = F\left(\tanh \frac{\vf}{\sqrt6 \a}\right) \,.
	\eea
	In this case, the simplest representative $V = \l_1 \tanh^{2} \frac{\vf}{\sqrt 6 \a}$ gives rise to the same inflationary predictions as the Starobinsky model \cite{Starobinsky}, the chaotic inflation with a quartic potential and non-minimal coupling \cite{Hig1,Hig2,Hig3,Hig4,Hig5,Hig6} and the supersymmetric completion of these models as well as a broad class of (super)conformal attractors \cite{SuperStraobinsky,Nonmin,Con1,Con2,Con3,Con4,Con5} for a vast range of $\a$. For $\a \rightarrow \infty$, the inflationary predictions become identical to the chaotic inflation with a quadratic potential\cite{alpha}
	\bea
	1- n_s = \frac{2}{N}\,, \qquad r = \frac{8}{N} \,.
	\label{Chaotic}
	\eea
	Thus, the model interpolates between the attractor point (\ref{Attractor}) and the chaotic inflation (\ref{Chaotic}).
	
	As the $\a$-attractors lead to inflationary predictions that perfectly fit the latest Planck data for a wide range of $\alpha$ \cite{KalloshCMB}, it is desired to understand the theoretical origin of the $\alpha$ parameter. There have been numerous attempts in this direction in recent years. For example, in the superconformal construction of the $\a$-attractors, the $\a$ parameter is a measure of the curvature of the inflaton K\"ahler manifold \cite{alpha}. This parameter also appears in other contexts such as the string field theory approach \cite{Stringinf}, spacetime with vector distortion \cite{Jimenez1}, and scale-invariant gravity \cite{OzkanRoest,Akarsu}. A particular approach, called the auxiliary vector modification \cite{OzkanTsu}, refers the $\alpha$ parameter as the Brans-Dicke parameter $\o_{BD}$ by replacing the Ricci scalar $R$ with an auxiliary vector modified Ricci scalar
	\bea
	R \quad \rightarrow \quad R + A^\m A_\m + \b \, \nabla^\m A_\m \,,
	\eea
	in $f(R)$ theories of gravity where $\b$ is a free parameter. In this class of theories, the vector field $A_\m$ is an auxiliary field, thereby such models are equivalent to a scalar-tensor theory with a particular scalar potential depending on the choice of the function $f(R)$. To show that, one first rewrites the auxiliary vector modified $f(R)$ theory as 
	\bea
	\cL &=& f(R + A^\m  A_\m + \b \nabla^\m A_\m) \nonumber\\
	&=&  f(F) - \vf \left( F - R - A^\m  A_\m -  \b \nabla^\m A_\m \right)\,.
	\eea
	Note here that the elimination of $\vf$ precisely recover the auxiliary vector  modified $f(R)$ theory. One may now eliminate $A_\m$ and perform a Weyl rescaling $ g_{\m\n} \rightarrow \vf g_{\m\n}$, which gives rise to the Einstein-frame action with a canonical scalar field $(\phi)$ with a scalar potential \cite{OzkanTsu}
	\begin{equation}
	V(\phi)  = \frac{M_{\textrm{pl}}^2}{2}  e^{-\sqrt{\frac{2}{3\alpha}} \frac{\phi}{M_{pl}} } \left( F -  e^{-\sqrt{\frac{2}{3\alpha}} \frac{\phi}{M_{pl}} } f(F)\right) \,,
	\label{GenericPot}
	\end{equation}
	where $\alpha$ is defined  in terms  of $\b$ and the Brans-Dicke parameter $\o_{BD}$ as
	\begin{equation}
	\alpha \equiv 1+\frac{\b^{2}}{6}=1+\frac{2}{3} \omega_{\mathrm{BD}}  \,.
	\end{equation}
	Here $f(F)$ satisfies $f_{,F} (F) = e^{\sqrt{\frac{2}{3\alpha}} \frac{\phi}{M_{pl}} }$. When this modification is considered for the auxiliary modified Starobinsky model $f(F) = F + (1/6M^2) F^2$, one obtains a scalar potential 
	\bea
	V(\phi)=\frac{3}{4} M_{\mathrm{pl}}^{2} M^{2}\left(1-e^{-\sqrt{\frac{2}{3 \alpha}} \frac{\phi}{M_{\mathrm{pl}}}}\right)^{2} \,.
	\eea
	This model clearly interpolates between the Starobinsky inflation (corresponding to $\a=1$) and the chaotic inflation with quadratic potential (corresponding to $\alpha \rightarrow \infty$), presenting a similar behavior as the $\alpha$-attractors \cite{alpha}.
	
	The generic form of the potential for the auxiliary vector modified $f(R)$ theories (\ref{GenericPot}) suggests a more general interpolating behavior than that of the Starobinsky and $\f^2$ chaotic inflation. In particular, if $f(F)$ is chosen to be \cite{OzkanTsu}
	\bea
	f(F) = F + m^{2(1-n)} F^n \,, 
	\eea
	then one ends up with a scalar potential
	\bea
	V(\phi)=\frac{(n-1)}{2 n^{n /(n-1)}}  m^{2} e^{-2 \sqrt{\frac{2}{3 \alpha}} \f}\left(e^{\sqrt{\frac{2}{3 \alpha}} \f}-1\right)^{\frac{n}{n-1}} \,, \quad
	\label{RmFnPotential}
	\eea
	where we set $M_{pl} = 1$. In this case, there is an interpolation between the $R + R^n$ type of inflationary models  (corresponding to $\a=1$) and the chaotic inflation with the power-law potential $V(\f) \propto \f^p$ (corresponding to $\alpha \rightarrow \infty$) where 
	\bea
	p \equiv \frac{n}{n-1} \,.
	\eea
	Consequently, this choice generalizes the interpolation between $R^2$ and $\phi^2$ of the $\alpha$-attractors to a more general interpolation between $R^n$ and $\phi^p$. Here, we refer to these models as $R^n - \phi^p$ interpolating inflationary models. It is important to note here that while generalizing the $\alpha$ attractors to a broader class of models, the generic $R + R^n$ models do not belong to any of the universality classes and they do not present an attractor point \cite{Encyc}. In that sense, the $n=2$ choice is still special due to its attractor behavior.
	
	In this paper, our purpose is to embed this generic $R + R^n$ and $\phi^p$ interpolating models to supergravity as a natural next step since supersymmetry is the leading proposal of physics beyond the Standard Model.  As the implementation of the $\alpha$-attractors follow the superconformal embedding by considering the couplings of chiral multiplets, here we will follow a similar path and first briefly discuss the elements of four dimensional $\cN=1$ conformal  supergravity and its chiral multiplet couplings in Section \ref{Section3}. Next, we discuss the superconformal embedding of $R^n - \f^p$ inflationary model, then we gauge fix the redundant (super)conformal symmetries. At the end of the Section \ref{Section3} we obtain the stability condition during inflation in terms of the scalar masses. Finally, we  discuss the phenomenology of the interpolating $R^n$ -- $\phi^p$ models. In particular, we will concentrate on the role of the $\alpha$ parameter in bringing the disfavored $R + R^n$ and $\phi^p$ models into the favored region of the latest Planck data. Furthermore, we will discuss the stability of a number of favored supersymmetric examples.  In Section \ref{Section4}, we give conclusion and discussions.

	\section{Superconformal Generalizations of $R^n - \f^p$  Interpolating  Inflationary Models}\label{Section3}
	
In this section, we describe the superconformal realization of a minimally coupled scalar-tensor theory with a scalar potential (\ref{RmFnPotential}). This can be achieved in a superconformal setting  by properly choosing the conformal K\"ahler potential $N(X,\bar X)$ that is a real function of Weyl weight 2 which satisfies 
		\bea
		N = N(X,\bar{X}) = X^I N_I = \bar{X}^I N_{\bar I} = X^I \bar{X}^{\bar J} N_{I \bar J} \,.
		\eea
	as well as a conformal superpotential $\cW = \cW(X)$ that is an arbitrary holomorphic function of $X$ and has Weyl weight 3 with the following property
	\bea
	\cW_I \equiv \frac{\partial \cW}{\partial X^I} \,, \qquad \text{with} \qquad  X^I \cW_I= 3 \cW   \,.
	\label{ConformalSuperpotential}
	\eea
	As the the interactions of scalars $X^I$ are determined by the  K\"ahler metric in a rigid theory, the function $N_{I  \bar J}$ can be understood in terms of K\"ahler metric 
\bea
G_{I\bar J} = N_{I  \bar J} \,,
\eea
derived from the superconformal K\"ahler potential $N(X,\bar X)$. Thus, the superconformal action is entirely determined by the metric $G_{I\bar J}$, its inverse and the superpotential $\cW$
\bea
e^{-1} \cL &=& - \frac{1}{6} N R- G_{I\bar{J}} D_\m X^I D^\m \bar{X}^{\bar J} - G^{I\bar{J}} \cW_I \bar{\cW}_{\bar J} \,, \quad
\label{physical}
\eea
Here, the covariant  derivative of $X^I$ is defined as
\bea
D_\m X^I = \left(\partial_{\mu} - \rmi A_{\mu}\right) X^I \,,
\eea
where  $A_\m$ appearing inside the covariant derivative is  given by
\bea
A_\m &=& \frac{\rmi}{2N} \left(N_{\bar I} \partial_\m \bar{X}^{\bar I} - N_{ I} \partial_\m {X}^{I}  \right) \,,
\eea
Armed with the superconformal action (\ref{physical}) we can proceed to the construction of the supersymmetric completion of a minimally coupled scalar-tensor theory with a scalar potential (\ref{RmFnPotential}). We achieve this goal by using three chiral
	multiplets. One of these multiplets is to be used to compensate for the superconformal symmetries, which we call a compensating multiplet with the fields $\{X^0, P_{L} \l^0, F^0\}$.  Another multiplet, which we will use to realize inflation is called an inflaton multiplet with the fields $\{X^1, P_{L} \l^1, F^1\}$. The real part of the scalar $X^1$ will play the role of the inflaton. Finally, we need a third multiplet whose consistent truncation will help us to realize the scalar potential (\ref{RmFnPotential}) in the inflationary paradigm. This multiplet is referred to as the Goldstino multiplet, which consists of the fields $\{S, P_{L} \p, M\}$ that is labeled with $I=2$. The compensating multiplet has no role in the inflationary dynamics but the scalar of the Goldstino multiplet $S$ and the imaginary part of the scalar of the inflaton multiplet $\textrm{Im}(X^1)$ could potentially cause serious problems for the stabilization of the inflationary trajectory. This problem was addressed in the context of $\alpha$-attractors for various choices of the K\"ahler potential and superpotential \cite{SuperStraobinsky,alpha, Marco1,Marco2,Marco3}. A more general analysis for the abovementioned superconformal setting with three chiral multiplets was given in \cite{KalloshRube} and \cite{Zavala1}, thus we will briefly review the results of \cite{KalloshRube} and \cite{Zavala1} before presenting various choices of superconformal K\"ahler and superpotentials that gives rise to a stable superconformal generalization of the scalar potential (\ref{RmFnPotential}).
	
	The defining property of the superconformal superpotential (\ref{ConformalSuperpotential}) implies that it can be solved as
	\bea
	\cW =  (X^0)^2  \,S \,f\left(\frac{X^1}{X^0}\right) \,,
	\eea
	where $f\left(\frac{X^1}{X^0}\right) $ is an arbitrary holomorphic function. After gauge fixing the dilatation symmetry by using $X^0$, the superconformal superpotential becomes the superpotential $W$ that is of the form
	\bea
	W = S\, f(X^1) \,.
	\eea
	Along the inflationary trajectory $S = \rm{Im}(X^1) = 0$, the generic form the scalar potential in the Einstein frame is given in terms of the K\"ahler and the superpotential as
	\bea
	V &=& e^K \left( K^{\a \bar{\b}} (\partial_\a + K_\a) W  (\partial_{\bar \b} +  K_{\bar \b}) \bar{W}  - \frac{3}{M_{pl}^2} |W|^2 \right)\nn\\
	&\approx&  e^K K^{S \bar{S}} |f|^2 \,,
	\eea
	since the superpotential only linearly depends on $S$. Here $\a, \b = 1,2$ due to the use of the compensating multiplet for gauge fixing. At this point, let us discuss two symmetry argument to be imposed on the K\"ahler potential, which will be essential in finding a supersymmetrization of the scalar potential (\ref{RmFnPotential}). 
	
	\begin{enumerate}
		\item {The K\"ahler potential can be chosen not to depend on the real part of $X^1$. In this case, the exponential factor $e^K$ as well as the inverse metric $K^{S\bar S}$ becomes irrelevant in determining the inflationary potential if one imposes the following symmetries \cite{KalloshRube}
			\bea
			&& S \rightarrow -S \,, \quad X^1 \rightarrow \bar{X}^1 \,,  \quad  X^1 \rightarrow X^1 + c \,.
			\label{KahlerCriteria}
			\eea 
			where $ c \in \mathbb{R}$. These conditions make sure that the inflationary potential is only determined  by the choice of $f$. Consequently,  at $S = 0$ and $\rm{Im}(X^1) = 0$,  the scalar potential takes a very simple form
			\bea
			V = f^2(\rm{Re}(X^1)) \,.
			\eea 
			For the stability of this inflationary trajectory $S = \rm{Im}(X^1) = 0$ against the  small fluctuations of the fields $S$ and $\rm{Im}(X^1)$ one needs to ensure that they are heavier than the Hubble scale $H$ \cite{Ferrara1,Ferrara2,KalloshRube}. If we decompose the scalar fields $X^1$ and $S$ into their real and imaginary parts as
			\bea
			X^1 = \frac{1}{\sqrt2} (\phi + \rmi \l)\,, \qquad S = \frac{1}{\sqrt2} (s + \rmi \g) \,,
			\eea
			the masses $m_s, m_\l$ and $m_\g$ read \cite{KalloshRube}
			\bea
			&& m_\l^2 \approx 6H^2 (1- K_{X^1 \bar{X^1} S \bar{S}}) \,, \nn\\
			&&  m_s^2 = m_\g^2 \approx - 3 H^2 K_{S \bar{S} S \bar{S}} \,.
			\label{mass}
			\eea
			during single field slow-roll inflation. Thus, the stability requires the K\"ahler potential to satisfy \cite{KalloshRube}
			\bea
			K_{X^1 \bar{X^1} S \bar{S}} \lesssim \frac{5}{6}\,, \qquad K_{S \bar{S} S \bar{S}}  \lesssim - \frac{1}{3} \,.
			\eea
			As a future reference, we emphasize that if a K\"ahler potential has a non-zero $K_{S \bar{S} S \bar{S}}$ derivative, it only contributes to the mass of $s$ and $\g$ but not the mass of $\l$. Thus, the stability of  $s$ and $\g$ can always be achieved by modifying the K\"ahler potential with $(S \bar S)^2$ term, in which case  $K_{S \bar{S} S \bar{S}}$ does not vanish during inflation.
		}
		\item{The criteria (\ref{KahlerCriteria}) only allows terms of type $X^1 - \bar X^1$ due to the shift symmetry of $X^1$. We may allow  $X^1 + \bar X^1$ type of terms in which case the scalar potential does depend on both the exponential factor $e^K$ as well as the inverse metric $K^{S \bar S}$. In this case, a possible  symmetry in the space of complex fields depends on the choice of the K\"ahler potential. A particular choice, which will be useful in our construction, is given  by
			\bea
			K = - \a \log (\Phi) \,, \quad \text{with} \quad \Phi = X^1 + \bar X^1 - S \bar S  \,.
			\eea
			This choice has the following symmetry amongst the fields \cite{Zavala1}
			\bea
			&& X^1 \rightarrow X^1 + \rmi a + \bar{b} S + \frac12 b \bar b\,, \quad S \rightarrow S + b   \,,
			\eea
			where $a \in {\mathbb R}\,, b \in {\mathbb C}$. With this choice of K\"ahler potential, the masses for the imaginary part of $X^1$ and $S$ read \cite{Zavala1}
			\bea
			m_{\l}^2 &=& \frac2\a \Big[  X^{1-\a} \left(1 - \frac1\a\right) f^2 - \frac{X^{2-\a}}{\a} f f^\prime \nn\\
			&& + \frac{X^{2-\a}}{2\a} \left( f^{\prime 2} - f f^{\prime \prime}\right)\Big] \,,\nn\\
			m_{\g}^2  = m_{s}^2 &=& \frac{X^{1-\a}}{\a} \left(\a - 2 - \frac1\a \right) f^2 \nn\\
			&& + \frac{X^{2-\a}}{\a} \left(\frac2\a - 2\right) f^\prime f + \frac{X^{3-\a}}{\a^2} f^{\prime 2} \,,
			\label{MassNonD}
			\eea
			where and the superpotential is taken as $W = S f(\Phi)$. }
	\end{enumerate}
	
	We now paved the way for the analysis of the supersymmetric models and their stability. As our first example, we introduce a model for a stable supersymmetrization of the bosonic scalar potential (\ref{RmFnPotential}) with a canonical kinetic term for the inflaton. This means that the K\"ahler potential does not consist of the interpolation parameter $\a$ and consequently the masses and the stability is $\alpha$-independent. To achieve that, let us consider the following superconformal K\"ahler potential and superpotential
	\bea
	\cW(X) &=& \frac13 \left( \frac{(n-1)m^2}{2 n^{\frac{n}{n-1}}} \right)^{\frac12} S (X^0)^2  e^{(-2 \sqrt{\frac{1}{\a}} \frac{X^1}{X^0})} \nn\\
	&& \times \left(  e^{ 2\sqrt{\frac{1}{\a}} \frac{X^1}{X^0}}  -1 \right)^{\frac{n}{2(n-1)}} \,,\nn\\
	N(X, \bar{X}) &=& - |X^0|^2  e^{\left( - \frac{|S|^2}{|X^0|^2}  + \frac12 \left( \frac{X^1}{X^0} - \frac{\bar{X^1}}{\bar X^0}  \right)^2  + 3 \xi \frac{|S|^4}{|X^0|^4} \right)} \,, \qquad 
	\eea
	where $\xi$ is a free parameter which will play an important role in the stability of the model. We impose the following gauge fixing conditions
	\bea
	X^0 = \sqrt3\,, \qquad b_\m = 0 \,, \qquad P_L \O^0 = 0 \,,
	\label{gauge}
	\eea
	where the first condition fixes dilatation while the second one fixes the special conformal symmeties and the last one fixes the $S$-SUSY. The corresponding K\"ahler and the superpotentials are then given by
	\bea
	W &=&\left( \frac{(n-1)m^2}{2 n^{\frac{n}{n-1}}} \right)^{\frac12} S e^{- 2\sqrt{\frac{1}{3 \a}}{X^1}} \nn\\
	&& \times  \left(  e^{2\sqrt{\frac{1}{3\a}} {X^1}}  -1 \right)^{\frac{n}{2(n-1)}}\,,\nn\\
	K  &=& S \bar{S} - \frac12 \left(X^1 - \bar{X^1} \right)^2 - \xi  |S \bar{S}|^2 \,.
	\eea
	Note that the K\"ahler potential is of the desired form (\ref{KahlerCriteria}). At $S= \rm{Im}(X^1) = 0$, the scalar potential is exactly given by (\ref{RmFnPotential}) as desired.
	In this case, the stability condition (\ref{KahlerCriteria}) becomes a simple criteria to be imposed on the parameter $\x$
	\bea
	\x \geq \frac{1}{12} \,, \label{masses1}
	\eea
	which is identical to the stability of the Starobinsky model for a certain choice of superpotential and K\"ahler potential \cite{SuperStraobinsky}.
	
	We may also choose to have a non-canonical kinetic term to introduce the exponential function in the scalar potential (\ref{RmFnPotential}) by redefining the inflaton field $\phi$. This can be most easily achieved by having a kinetic term of type $(1/\f^2)(\partial \f)^2$ which can be achieved by the following choice of superconformal K\"ahler potential
	\bea
	N (X, \bar{X}) &=& - |X^0|^2  \Bigg(   \sqrt{3}\left(\frac{X^1}{X^0} + \frac{\bar{X^1}}{\bar X^0}\right) - \frac{3 S \bar{S}}{|X^0|^2} \nn\\
	&& \qquad  \qquad + 9 \zeta \frac{(S \bar{S})^2}{|X^0|^4} \Bigg)^\a  \,.
	\eea
	Then choosing  the superconformal superpotential to be
	\bea
	\cW (X) &=& \frac{1}{\sqrt3}\left( \frac{\a (n-1) m^2}{2^{2-3\a} n^{n/(n-1)}}\right)^{\frac12}  (X^0)^2 S \nn\\
	&& \times \left(\frac{\sqrt 3 X^1}{X^0} - 1\right)^{\frac{n}{2(n-1)}} ( \frac{\sqrt 3 X^1}{X^0} )^{\frac{3 (\a-1)}{2}} \,,
	\eea
	in accordance with our choice of the  superconformal K\"ahler potential, we achieve  the following superpotential and the K\"ahler potential after gauge fixing (\ref{gauge})
	\bea
	W &=& \left( \frac{3\a (n-1)}{2^{2-3\a} n^{n/(n-1)}}\right)^{\frac12} m S \nn\\
	&& \times \left(X^1 - 1\right)^{\frac{n}{2(n-1)}} (X^1)^{\frac{3 (\a-1)}{2}}\,,\nn\\
	K  &=& -3 \alpha \log \left( X^1 + \bar{X^1} - S \bar{S} + \zeta \frac{(S \bar{S})^2}{X^1 + \bar X^1} \right) \,.
	\eea
	Once again, at  $S= \rm{Im}(X^1) = 0$, the bosonic part of the supersymmetric theory precisely generates the desired scalar potential (\ref{RmFnPotential}) with the slight change of notation
	\bea
X^1 = (\phi + \rmi \l)\,, \qquad S = (s + \rmi \g) \,.
\eea
The notational change was necessary at this point to be consistent with the notations of \cite{Zavala1} and consequently to read off the masses from (\ref{MassNonD}). At the inflationary trajectory $S = \rm{Im}(X^1) = 0$, these  masses can read off using (\ref{MassNonD}) and are given by
	\bea
	m_\g^2 = m_{s}^2  &=&  \frac{1}{\a} \left( -2 + 4 \x\right) H^2 \,, \nn\\
	m_\l^2 &=& \frac{2H^2 (3\a - 1)}{\a} \,. \label{masses2}
	\eea
	Note that $\xi$ term only modifies the mass of  $\g$ and $s$ since the additional term always vanishes at $S=0$. These degrees of freedom must be stabilized around (or above) the Hubble scale  which can always be achieved by appropriate choice of $\xi$ as long as $\a > \frac13$. 
	\begin{figure}
		\centering
		\includegraphics[scale=0.2]{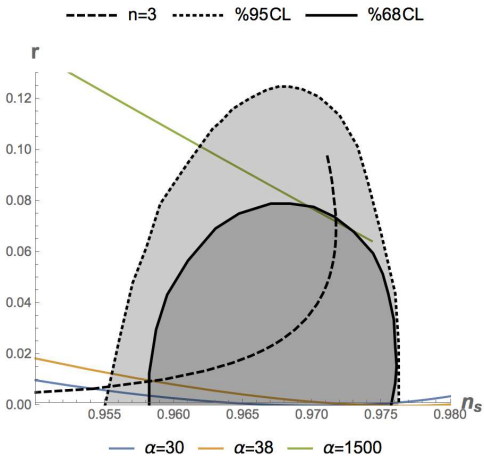}
		\caption{The theoretical curves for the $n=3$ theory in the spectral index ($n_s$) versus the tensor-to-scalar ratio ($r$) plane. The dashed line $n=3$ represents the values of $n_s$ and $r$ for $N=60$ for each $\a$. Around $\a=30$, this trajectory enters to the $ \% 95 $ Confidence Level (CL) and between $\a =38-1500$ it remains inside the $ \%68$ CL.}
		\label{Fig1}
	\end{figure}
	\begin{figure}
		\centering
		\includegraphics[scale=0.25]{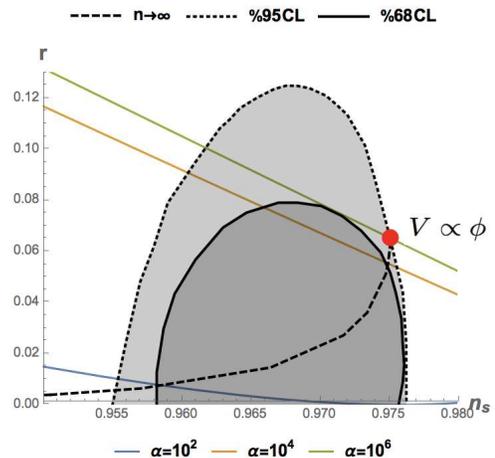}
		\caption{In the $n\gg 1$ limit, the model approximates to $V \propto \f$ when $\a \rightarrow \infty$ that is marginally inside the $\%95$ CL for $N=60$. Around $\a = 10^2$, the $n\rightarrow\infty$ trajectory enters to the $ \%68 $ CL and remains inside until $\a \approx 10^4$}
		\label{Fig2}
	\end{figure}
	
	We would like to elaborate this point by discussing the phenomenology of supersymmetric $R^n - \f^p$ interpolating models in terms of tensor to scalar ratio $r$  and spectral index $n_s$ while demanding that the $s,\g$ and $\l$ are stabilized around (or above) the Hubble scale. Since the inflationary observables would presumably depend on $\a$, they should provide constraints on the stabilizing term $\xi$ which will allow us to see if the stability of the supersymmetric models are compatible with the constraints on $\a$ from the Planck data. The case of  $n=2$ was already examined in \cite{alpha} where $\x$ was taken as $\x = 3g$. Therefore, we will focus on the analysis of  $n>2$, in particular $n=3,4,5$. For $n=3$, this model simply reduce to  $R + R^3$ for  $\a=1$ which is outside of the Planck contour \cite{Encyc}. Increasing $\alpha$ pushes the model into regime $50 \lesssim N \lesssim 60$ which is viable region of Planck data \cite{Planck2018}, see Fig.\ref{Fig1}. For $\a=30$, the model is marginally inside the $\% 95$ CL and between $\a =38-1500$, model is inside the $\%68$ CL. Within this region, the stability can be achieved by demanding i.e. $\xi > 400$. For $n=4$, the inflationary observables are inside the $\%68$ CL for  $\a =55-3000$ and the stability can be achieved by choosing, i.e. $\x > 750$. For $n=5$, we need to choose $\a = 65 - 5000$ to stay inside the $\%68$ CL and the stability can be achieved by choosing, i.e. $\x > 1250$. Finally, we note  that for $n \gg 1$, the scalar potential is given by
	\bea
	V(\f) \approx \frac12 m^2 e^{-2\sqrt{\frac{2}{3\a}}\f} \left(- 1 + e^{\sqrt{\frac{2}{3\a}}\f}  \right) \,,
	\eea
	which approximates to a linear potential in $\a \rightarrow \infty$ that in marginally inside the $\%95$ CL. On the other hand, for $\a \approx 10^{2} - 10^4$ the inflationary observables stay inside the $\%68$ CL., see Fig.\ref{Fig2}. For this range, the stability of the model is satisfies as long as $\x > 2500$.

	\section{Conclusions}\label{Section4}
	In this paper, we present examples of superconformal and supersymmetric completion of the $R^n - \f^p$ inflationary models. These models interpolate between the polynomial $f(R)$ theories and chaotic inflation with a power-law potential. They are in the $\%68$ CL for a certain range of  $\a$ and are stable under certain restriction on $\x$, which becomes more restrictive as $n$ is chosen large.
	
	The supersymmetrization procedure we utilized here makes use of three chiral multiplets. One of them, which we referred to as the Goldstino multiplet, is particularly important in the supersymmetric realization of the inflationary models as it allows a trajectory for successful inflation.  Perhaps a better way to consider a supersymmetric completion could be the direct supersymmetrization of the Lagrangian
	\bea
	e^{-1} \cL &=& \widetilde R + m^{2(1-n)}  \widetilde R^n \,,
	\eea
	where $\widetilde R = R + A^\m  A_\m + \b \nabla^\m A_\m$. Alternatively, it would be interesting to achieve the same supersymmetrization with less number of fields, i.e. with only the conformon and the inflaton supermultiplets. Finally, the non-canonical scenario that we realized here constraints $\a$ with a lower bound $\a > \frac13$. It would be interesting to find stable scenarios that allow one to smoothly go to $\a=0$, allowing interpolation between $\a=0$ and  $\a  \rightarrow \infty$. 
	
	\section{Acknowledgments}
	
	The work of SB, EOK and MO is supported in part by TUBITAK grant 117F102. NO and UZ are supported in part by Istanbul Technical University Research Fund under grant number TDK-2018-41133.

\end{document}